\documentclass[%
reprint,
superscriptaddress,
groupedaddress,
nobibnotes,
 amsmath,amssymb,
 aps,
prl,
showkeys
]{revtex4-2}
\usepackage[dvipsnames]{xcolor}
\usepackage{graphicx}
\usepackage{dcolumn}
\usepackage{bm}
\usepackage[breaklinks=true,colorlinks,citecolor=blue,linkcolor=red,urlcolor=blue]{hyperref}

\usepackage{braket}
\usepackage[utf8]{inputenc} 
\usepackage{mathtools}
\usepackage[bottom]{footmisc}
\usepackage{hyperref}
\usepackage{amsfonts}
\usepackage{amsmath}
\usepackage{slashed}
\usepackage{amsthm}

\usepackage{tikz}

\usepackage{color}
\definecolor{mycolor2}{rgb}{1.0, 0.0, 0.0}
 
\begin{document}
\title{Quantum Geometry Driven Finite-Momentum Exciton Fluctuations in Flat-Band Systems}
\author{Yi-Chun Hung}
\email[Contact author: ]{hung.yi@northeastern.edu}

\author{Xiaoting Zhou}

\author{Arun Bansil}
\email[Contact author: ]{ar.bansil@northeastern.edu}

\affiliation{Department of Physics,\;Northeastern\;University,\;Boston,\;Massachusetts\;02115,\;USA}
\affiliation{Quantum Materials and Sensing Institute,\;Northeastern University,\;Burlington,\;Massachusetts\;01803,\;USA}
\date{\today}

\begin{abstract}
Quantum geometry is instrumental in stabilizing exotic phenomena in systems ranging from topological insulators to superconductors. In dispersionless flat bands, where the kinetic energy is quenched, the quantum metric emerges as the fundamental driver of macroscopic collective phenomena. Here, we theoretically demonstrate that lattice-geometry-induced flat bands, such as those in kagome and Lieb lattices, provide a fertile platform for realizing a purely quantum-geometry-driven excitonic insulator (EI) phase. By applying an out-of-plane Zeeman field to lift spin degeneracy without spin-orbit coupling, we establish a Ginzburg-Landau framework in which the electron-hole wavefunction-overlap directly maps the flat-band quantum metric onto the macroscopic free energy. This mapping plays a key role in both the EI and the associated superfluid phases, with the coherence length and phase stiffness emerging directly from the quantum metric. Our analysis reveals that under strong interactions, the quantum metric induces a negative effective kinetic coefficient for the amplitude mode. Rather than destabilizing the uniform condensate, this softens the amplitude fluctuations at a finite momentum, giving rise to a finite-momentum superfluid density fluctuation (FMSDF) state. This state is observable as a periodically modulated magnitude of in-plane magnetization fluctuations. Our findings establish a rigorous link between flat-band quantum geometry and dynamic collective excitonic states, with promising pathways for realization in covalent-organic frameworks (COFs).

\end{abstract}


\maketitle

\par The concept of quantum geometry is rapidly reshaping our understanding of materials and their responses, offering profound insights into topological insulators, unconventional superconductors, and correlated electronic systems. Originating from the overlap of Bloch wave functions across the Brillouin zone \cite{PhysRevB.56.12847, RevModPhys.84.1419}, quantum geometry is characterized by the complex quantum-geometric tensor. Its imaginary part, the Berry curvature, governs topological responses, while its real part (quantum metric) measures the distance between quantum states, playing a pivotal role in optical transitions and transport phenomena \cite{PhysRevX.14.011052, PhysRevLett.133.206602, PhysRevResearch.7.023158}. The quantum metric becomes a dominant feature in flat-band systems: As the kinetic energy is quenched, most phenomena are bounded at a fundamental level by the quantum metric \cite{PhysRevResearch.5.L012015, PhysRevX.14.041004}. Notable examples include the quantum-metric origin of superfluid weight and coherence length in flat-band superconductors \cite{PhysRevB.95.024515, PhysRevB.96.064511, Peotta2015, PhysRevLett.124.167002, PhysRevLett.117.045303, PhysRevLett.128.087002, Yu2025, PhysRevLett.132.026002, li2025vortex}, as well as the emergent spin stiffness in flat-band magnetism \cite{PhysRevB.102.165118}.

\par Beyond superconductivity and magnetism, flat-band excitons present an intriguing frontier \cite{Tran2019, Yuan2020, Jiang2021, Jin2019, Shimazaki2020, Wilson2021}. Driven by both experimental \cite{PhysRevLett.116.066402, PhysRevLett.120.097401, PhysRevA.104.063505, pub.1059135791} and theoretical \cite{PhysRevB.69.085325, PhysRevLett.126.196403, PhysRevLett.130.186401, PhysRevLett.132.236001, PhysRevB.105.L140506, PhysRevB.101.195310, ying2024flatbandexcitonsquantum, doi:10.1073/pnas.2401644121} advances, recent studies on heterostructures have illustrated how the quantum metric influences the excitonic effective mass \cite{ying2024flatbandexcitonsquantum, Phys.Rev.B112.014518}. While connections between exciton vorticity and Chern number differences have been identified \cite{doi:10.1073/pnas.2401644121}, the precise mechanistic role of the quantum metric in determining physical observables, such as superfluid phase stiffness and coherence length, remains an open fundamental question. 

\par Flat bands arising from intrinsic lattice geometry, such as those found in kagome \cite{10.1143/ptp/6.3.306, PhysRevB.99.125131, PhysRevB.101.045131}, Lieb \cite{PhysRevLett.62.1201, PhysRevLett.117.045303, Slot2017, PhysRevLett.114.245504}, and dice lattices \cite{PhysRevB.34.5208, PhysRevB.108.075166, PhysRevB.108.075167}, offer exceptional simplicity and versatility \cite{Calugaru2022, PhysRevB.108.195140, Neves2024}. These flat bands can be effectively isolated from dispersive bands via mechanisms such as dimerization or anisotropic strain \cite{PhysRevB.106.155417, PhysRevB.99.125131, PhysRevB.101.045131, PhysRevB.99.155124, sharma2025straininducedtopologicalphase}, allowing enhanced control and tunability. Furthermore, recent studies demonstrate that twisted bilayer architectures based on these lattices yield isolated flat bands with highly tunable multiplicities \cite{PhysRevLett.133.236401, PhysRevB.109.155159, dice-on-graphene}. This exceptional controllability establishes these geometric frameworks as highly adaptable platforms for quantum material design. Motivated by this architectural flexibility, we investigate the formation and geometric stabilization of strongly correlated flat-band excitons within these systems.

\par
\begin{figure}[t]
  \centering
  \centering
    \includegraphics[width=\linewidth]{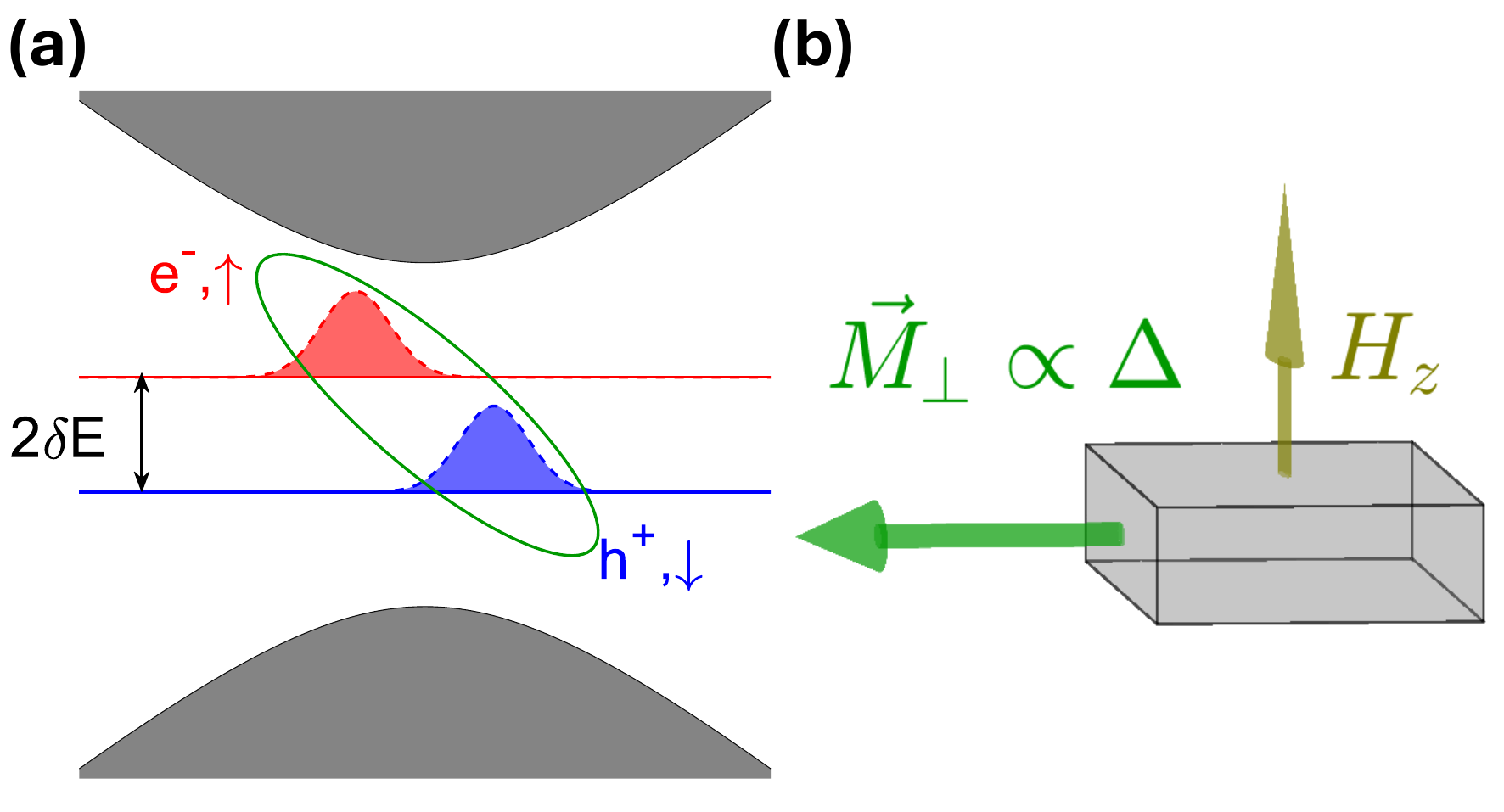}
  \caption{ Schematic diagrams for (a) the electronic structure of the flat band in our setup and (b) the exciton-induced magnetic response. $H_z$ is the out-of-plane Zeeman field with the corresponding energy splitting $2\delta E$.  $\vec{M}_\perp$ is the in-plane magnetization, which is proportional to the order parameter of the EI phase, $\Delta$. }
  \label{fig:01}
\end{figure}

 In this work, we demonstrate the emergence of a highly anomalous excitonic insulator (EI) phase driven exclusively by the quantum geometry of the lattice-induced flat bands. By introducing an out-of-plane Zeeman field to a system lacking spin-orbit coupling (SOC), we establish a purely spin-split valence and conduction flat-band system. The electron-hole wavefunction-overlap rigorously allows us to map the quantum metric into the Ginzburg-Landau (GL) free energy. We find that the quantum metric acts not merely as a stabilizer for the EI state and its Berezinskii-Kosterlitz-Thouless (BKT) transition, but it radically alters the fluctuation spectrum. Under strong interaction, the quantum metric drives a negative effective kinetic coefficient for the amplitude mode. Rather than collapsing the homogeneous condensate, this shifts the softest amplitude fluctuations to a finite momentum, leading to a state we define as the finite-momentum superfluid density fluctuation (FMSDF) phase. Furthermore, we establish a direct physical mapping between this EI phase and in-plane magnetization, identifying the FMSDF state as a periodic modulation of the magnetic fluctuations (Fig.~\ref{fig:01}(b)) observable in real materials such as covalent-organic frameworks (COFs) \cite{doi:10.1021/acs.accounts.0c00652, Jiang2019, D3SC06367D}.

\paragraph{Effective Hamiltonian and Geometry---} We consider a tight-binding model lacking SOC that naturally hosts a zero-energy flat band. An out-of-plane Zeeman field explicitly breaks the spin-$SU(2)$ symmetry, creating highly localized, spin-polarized conduction ($\uparrow$) and valence ($\downarrow$) flat bands separated by an energy gap $2\delta E$ (Fig.~\ref{fig:01}(a)). Assuming these flat bands are sufficiently isolated from the dispersive bulk bands, we project the system into the low-energy flat-band subspace. The coarse-grained effective Hamiltonian reads $H = H_0 + H_{\text{int}}$, where
\begin{equation}\label{eq:02}
    H_0 = \int d[\mathbf{r}] \, \delta E \big[ \hat{\psi}^\dagger_{\uparrow}(\mathbf{r})\hat{\psi}_{\uparrow}(\mathbf{r})-\hat{\psi}^\dagger_{\downarrow}(\mathbf{r})\hat{\psi}_{\downarrow}(\mathbf{r}) \big]
\end{equation}
is the single-particle Hamiltonian, and
\begin{equation}\label{eq:03}
    H_{\text{int}}=U_{c}\int d[\mathbf{r}] \, \hat{\psi}_{\uparrow}^\dagger(\mathbf{r})\hat{\psi}_{\downarrow}^\dagger(\mathbf{r})\hat{\psi}_{\downarrow}(\mathbf{r})\hat{\psi}_{\uparrow}(\mathbf{r})
\end{equation}
is the coarse-grained interaction between flat bands with different spin species, which mimics the Coulomb interaction in the material. Here, the field operators are defined by the Bloch states of the target flat band: $\hat{\psi}_{\sigma}(\mathbf{r}) = \int d[\mathbf{k}] \varphi_{\mathbf{k}}(\mathbf{r})\hat{\gamma}_{\sigma,\mathbf{k}}$, where $\varphi_{\mathbf{k}}(\mathbf{r})$ is the Bloch wavefunction of the target flat band and $\hat{\gamma}_{\sigma,\mathbf{k}}$ annihilates an electron with spin $\sigma$ and momentum $\mathbf{k}$ in that band.  Because SOC is absent, the Bloch wavefunctions $\varphi_{\mathbf{k}}(\mathbf{r})$ are strictly identical for both spin species. $U_{c}$ and $\delta E$ represent the coarse-grained interaction strength and energy separation between different spin species.  The chemical potential is set at zero for simplicity. The integral measures are $d[\mathbf{r}]=d^2\mathbf{r}$ and $d[\mathbf{k}]=d^2\mathbf{k}/(2\pi)^2$.

\par We introduce a path integral formulation for addressing the possible EI phase formed when the interaction is sufficiently large \cite{PhysRevB.107.155142}. Through a standard Hubbard-Stratonovich (HS) transformation, we obtain the partition function involving electrons and the order parameter (OP) for the EI phase:
\begin{equation}\label{eq:04}
    \mathcal{Z}=\int\mathcal{D}[\psi,\bar{\psi},\Delta,\Delta^*]e^{-S_0[\psi,\bar{\psi}]-S_{\text{mix}}[\psi,\bar{\psi},\Delta,\Delta^*]-S_{\Delta}[\Delta,\Delta^*]}.
\end{equation}
Here, $\psi=(\hat{\psi}_{\uparrow},\, \hat{\psi}_{\downarrow})^T$ and $\bar{\psi}=(\hat{\psi}_{\uparrow}^\dagger,\, \hat{\psi}_{\downarrow}^\dagger)^T$ denote two-component Grassmannian fields. The OP $\Delta$ satisfies the relation $\langle\Delta(\mathbf{r})\rangle=U_{c}\langle\hat{b}(\mathbf{r})\rangle$, where $\hat{b}(\mathbf{r})\equiv\hat{\psi}_{\downarrow}^\dagger(\mathbf{r})\hat{\psi}_{\uparrow}(\mathbf{r})$. The flat band's quantum geometry enters the action via
\begin{equation}\label{eq:08}
S_{\text{mix}} = \int^\beta_0 d\tau \int d[\mathbf{k}]d[\mathbf{q}]\Lambda(\mathbf{q},-\mathbf{k})\Delta_\mathbf{k}\hat{\gamma}^\dagger_{\uparrow, \mathbf{k}+\mathbf{q}}\hat{\gamma}_{\downarrow,\mathbf{q}}+ \text{H.c.}
\end{equation}
Here, $\Lambda(\mathbf{k},\mathbf{q}) = \braket{u_\mathbf{k-q}|u_\mathbf{k}}$ is the form factor, where $u_\mathbf{k}$ is the cell-periodic part of the flat band's Bloch wave function; see Supplemental Material (SM) for details \cite{SM}\nocite{grosso2013solid}. This form factor represents the electron-hole wavefunction overlap and relates to the quantum metric of the flat band through long-wave-length modes of which
\begin{equation}\label{eq:overlap}
\lim_{\mathbf{q}\to0}|\Lambda(\mathbf{k},-\mathbf{q})|^2\cong1-g_{ij}(\mathbf{k})\mathbf{q}^i\mathbf{q}^j,
\end{equation}
where $g_{ij}(\mathbf{k})=\text{Re}[\braket{\partial_{k^i}u_{\mathbf{k}}|(1-\ket{u_{\mathbf{k}}}\bra{u_{\mathbf{k}}})|\partial_{k^j}u_{\mathbf{k}}}]$ is the quantum metric \cite{PhysRevX.14.011052, PhysRevLett.133.206602, PhysRevResearch.7.023158} and the summation convention is applied. The quantum geometry of flat bands enters the effective action for the OP via Eq.~\eqref{eq:08}, impacting the EI phase as discussed next.

\par The OP's effective action arises by integrating out the Grassmann fields in Eq.~\eqref{eq:04} \cite{fluctuations_in_SC, altland2010condensed, Coleman_2015}:
\begin{equation}\label{eq:action_eff}
    S_{\text{eff}}[\Delta,\Delta^*]=\sum_{\mathit{q}}a(\mathit{q})\Delta_{\mathit{q}}^*\Delta_{\mathit{q}}+\sum_{\{\mathit{q}_n\}}b(\{\mathit{q}_n\})\Delta_{\mathit{q}_1}^*\Delta_{\mathit{q}_2}^*\Delta_{\mathit{q}_3}\Delta_{\mathit{q}_4}.
\end{equation}
Here, $\mathit{q}=(\mathbf{q},i\omega_n)$ and $\omega_n$ is the bosonic Matsubara frequency and $\sum_{\mathit{q}}=\sum_{\omega_n}\int d[\mathbf{q}]$; see SM for details \cite{SM}. For the long-wave-length mode in the static limit such that $\mathbf{q},\omega\to0$ \cite{PhysRevB.107.155142}, the quadratic coefficient is
\begin{equation}\label{eq:09}
a(\mathit{q}) \cong \frac{\sinh{\beta\delta E}}{1+\cosh{\beta\delta E}}\frac{K_{ij}}{4\pi\delta E}\mathbf{q}^i\mathbf{q}^j + a(0).
\end{equation}
Here, $\omega$ is the frequency obtained from analytic continuation of $i\omega_n\to\omega+i0^+$, $\mathbf{q}^i$ is the $i$th component of $\mathbf{q}$, and $K_{ij}=2\pi\int d[\mathbf{k}]g_{ij}(\mathbf{k})$. Since $K_{ij}$ is positive semi-definite \cite{PhysRevB.90.165139, PhysRevB.95.024515, Peotta2015, PhysRevX.14.011052, PhysRevX.14.011052, PhysRevLett.133.206602, PhysRevResearch.7.023158}, $\Delta_{\mathit{q}=0}$ is a stable ground state for flat bands with non-zero quantum metrics. Further, Eq.~\eqref{eq:09} shows that the coherence length $\xi_{\text{c}}$ is proportional to the quantum weight, once the mean-field order is formed; see SM for details \cite{SM}:
\begin{equation}\label{eq:xi_c}
    \xi_{\text{c}}^2 = \frac{K_{0}}{4\pi \Omega_{\text{BZ}}}|\frac{2\delta E(1+\cosh{\beta\delta E})}{g_{c}\sinh{\beta\delta E}} -  1|^{-1}
\end{equation}
where $\Omega_{\text{BZ}}$ is the area of the Brillouin Zone (BZ) and $g_{c}\equiv \Omega_{\text{BZ}}U_{c}$. Note that we assume an isotropic system such that $K_{ij}=\frac{K_0}{2}\delta_{ij}$ and $\text{tr}(K)=K_0$ without loss of generality.

\begin{figure}[t]
  \centering
  \centering
    \includegraphics[width=\linewidth]{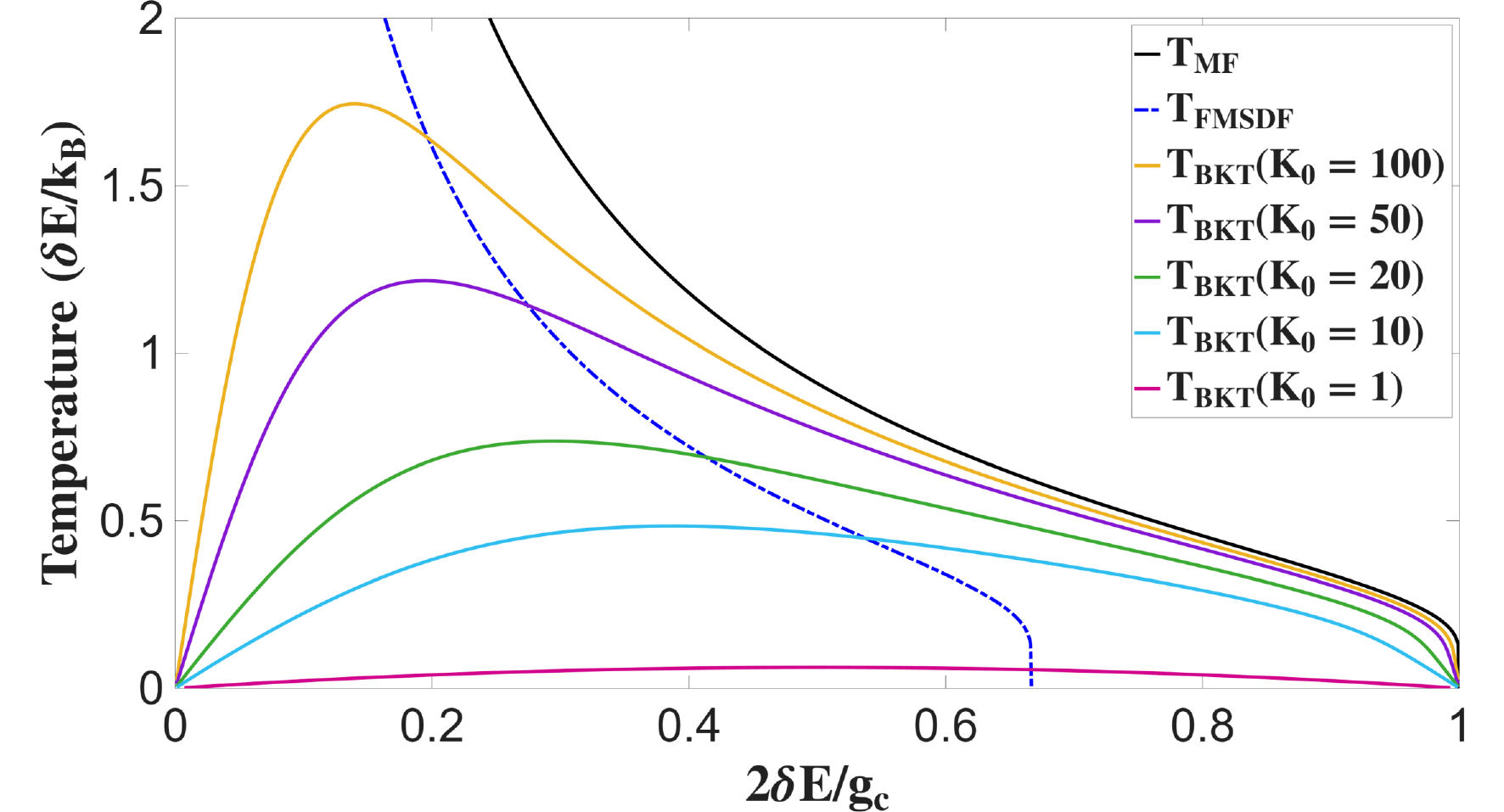}
  \caption{ Mean-field critical temperature $T_{\text{MF}}$ (black solid line), and the critical temperature for finite-momentum superfluid density fluctuation $T_{\text{FMSDF}}$ (black dashed line), and the BKT transition temperatures $T_{\text{BKT}}$ for different values of $K_0$ (colored lines) in the unit of $\delta E/k_B$ as a function of the dimensionless interaction strength $2\delta E/g_{c}$.
  }
  \label{fig:02}
\end{figure}

\paragraph{Mean-field theory and fluctuations---} Provided the quantum geometry of the flat bands stabilizes $\Delta_{\mathit{q}=0}$ as the ground state, we can expand the OP into its mean-field value $\Delta_0$ and the fluctuation $\delta \Delta_{\mathit{q}}$ near $\mathit{q}=0$ \cite{PhysRevB.107.155142}:
\begin{equation}\label{eq:10}
    \Delta_{\mathit{q}} = \sqrt{\beta}\Delta_0\delta_{\mathit{q},0} + \delta\Delta_{\mathit{q}}.
\end{equation}
We further define the corresponding Ginzburg-Landau (GL) free energy to analyze the phase transitions in our system \cite{PhysRevB.107.155142, fluctuations_in_SC, altland2010condensed}. The mean-field contribution of the GL free energy is:
\begin{equation}
    F_0 = \frac{1}{\beta}\bigg( a_0|\Delta_0|^2 + b_0|\Delta_0|^4 \bigg),
\end{equation}
where 
\begin{align}
    a_0 & = \frac{\beta \Omega_{\text{BZ}}}{2\delta E}(\frac{2\delta E}{g_{c}} - \frac{\sinh{\beta\delta E}}{1+\cosh{\beta\delta E}}), \label{eq:13}
    \\ b_0 & = \frac{\beta \Omega_{\text{BZ}}}{8(\delta E)^3}\frac{\sinh{\beta\delta E}}{(1+\cosh{\beta\delta E})}. \label{eq:14}
\end{align}
Based on Eq.~\eqref{eq:13}, a strong repulsive interaction within the flat bands is needed to exceed the band-gap energy (i.e., $g_{c}>2\delta E$) to facilitate a phase transition. Further, since $b_0$ is always positive according to Eq.~\eqref{eq:14}, such a phase transition is of second order. Once the phase transition occurs such that $a_0<0$, the amplitude of the OP can then be determined by the saddle point equation, resulting in $|\Delta_0|^2 = -a_0/2b_0$. Figure \ref{fig:02} shows the mean-field critical temperature $T_{\text{MF}}$ in the unit of $\delta E/k_B$ as a function of the dimensionless interaction strength $2\delta E/g_{c}$.

After analyzing the mean-field action, we focus on the OP fluctuations $\delta\Delta_{\mathit{q}}$ near $\mathit{q}=0$. The corresponding GL free energy is $F_{\delta}$, which includes terms up to the order $\mathcal{O}(|\delta\Delta|^2)$ and equals
\begin{equation}\label{eq:16}
    F_{\delta} = \frac{1}{\beta}\int d[\mathbf{q}] \sum_{\zeta=\pm}\lambda_{\zeta}(\mathbf{q}, |\Delta_0|)\eta_{\zeta,\mathbf{q}}^*\eta_{\zeta,\mathbf{q}}.
\end{equation}
Here, $\eta_{\pm,\mathbf{q}}= \sqrt{\beta}( e^{i\theta_0}\delta\Delta_{\mathit{q}} \pm e^{-i\theta_0}\delta\Delta^*_{-\mathit{q}} )/2$ with their corresponding eigenvalue $\lambda_{\pm}(\mathbf{q}, |\Delta_0|)$, where $\theta_0=\text{Arg}(\Delta_0)$. These modes correspond to the phase ($\eta_{-,\mathbf{q}}$) and amplitude ($\eta_{+,\mathbf{q}}$) fluctuations, respectively \cite{nagaosa1999quantum}. If we substitute the value for $|\Delta_0|=\sqrt{-a_0/2b_0}$ and consider the long-wave-length fluctuations, $\lambda_{\pm}(\mathbf{q})$ become
\begin{align}
    \lambda_{-}(\mathbf{q}) & \cong \frac{\beta }{g_{c}}\frac{K_{ij}}{\pi}\mathbf{q}^i\mathbf{q}^j, \label{eq:17}
    \\ \lambda_{+}(\mathbf{q}) & \cong -4a_0 + \frac{3\beta}{\delta E} \bigg(\frac{2\delta E}{g_{c}} - \frac{2}{3}\frac{\sinh{\beta\delta E}}{1+\cosh{\beta\delta E}} \bigg) \frac{K_{ij}}{2\pi}\mathbf{q}^i\mathbf{q}^j, \label{eq:18}
\end{align}
indicating that the phase mode is stable near the ground state and develops a superfluid phase; see SM for details \cite{SM}. According to Eqs.~\eqref{eq:17} and \eqref{eq:18}, the quantum metric of the flat bands plays a key role in the kinetics of the OP fluctuations, which further impact the SF and finite-momentum superfluid density fluctuation phases as discussed below. 

\paragraph{Superfluid (SF) and finite-momentum superfluid density fluctuation (FMSDF)---} According to Eq.~\eqref{eq:17}, the phase mode forms a superfluid (SF) phase whose phase stiffness is proportional to the quantum weight, consistent with the results of Ref.~\cite{Phys.Rev.B112.014518}. This indicates that the Berezinskii–Kosterlitz–Thouless (BKT) transition completely originates from the quantum metric of the flat bands, a feature that also appears in flat-band superconductivity \cite{PhysRevB.95.024515, PhysRevB.96.064511, Peotta2015, PhysRevLett.124.167002, PhysRevLett.117.045303, PhysRevLett.128.087002, Yu2025}. The corresponding BKT transition temperature $T_{\text{BKT}}=1/(k_B\beta_{\text{BKT}})$ satisfies the relation \cite{JMKosterlitz_1973, PhysRevLett.39.1201}:
\begin{equation}\label{eq:19}
    \frac{1}{\beta_{\text{BKT}}} = \frac{\pi}{2}|\Delta_{0,\text{BKT}}|^2\frac{1}{g_{c}}\frac{K_0}{2\pi},
\end{equation}
where $\Delta_{0,\text{BKT}}$ gives the value of $\Delta_{0}$ when temperature is $T_{\text{BKT}}$, and we assume an isotropic system such that $K_{ij}=\frac{K_0}{2}\delta_{ij}$ and $\text{tr}(K)=K_0$ without loss of generality. The BKT transition temperature $T_{\text{BKT}}$ in units of $\delta E/k_B$ as a function of the dimensionless interaction strength $2\delta E/g_{c}$ with different values of $K_0$ can be obtained by solving Eq.~\eqref{eq:19} self-consistently, as shown in Fig.~\ref{fig:02}; see SM for details \cite{SM}. The observed hump pattern in $T_{\text{BKT}}$ as a function of the interaction strength is attributed to the disruptions of the superfluid phase, which occur in cases of both very strong and very weak interactions, resulting in predominant Mott physics and hindering the condensation of excitons.

\begin{figure}[t]
  \centering
  \centering
    \includegraphics[width=\linewidth]{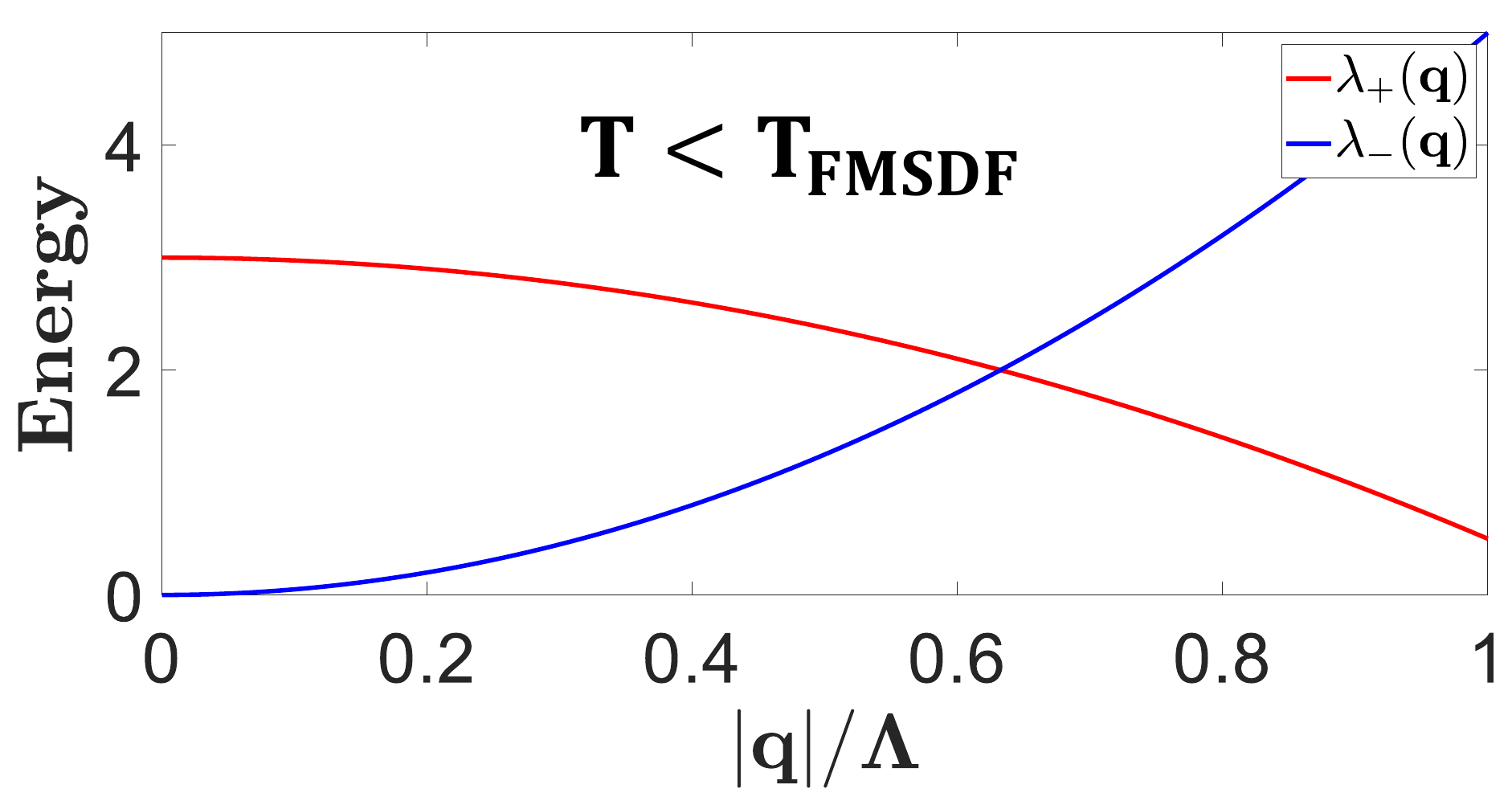}
  \caption{ A schematic diagram showing the energies of amplitude $\big(\lambda_+(\mathbf{q})\big)$ and phase $\big(\lambda_-(\mathbf{q})\big)$ fluctuations in the finite-momentum superfluid density fluctuation state, where $T<T_{\text{FMSDF}}$. Here, $\mathbf{\Lambda}$ is the momentum cutoff. The negative kinetic coefficient favors amplitude fluctuations with a finite momentum. However, the ground state remains as $\lambda_+(\mathbf{q})>0$ within the cutoff.
  }
  \label{fig:03}
\end{figure}

\par According to Eq.~\eqref{eq:18}, a strong interaction strength where $2\delta E/g_{c} < 2/3$ indicates a negative kinetic coefficient of the amplitude mode at finite temperatures, which implies that the amplitude mode favors components with finite momentum. Such a state does not break the translational symmetry of the ground state for a finite interaction strength since $\lambda_{+}(\mathbf{q})\geq2\beta/g_{c}>0$ within the momentum cutoff; see SM for details \cite{SM}. Therefore, we dub this as a finite-momentum superfluid density fluctuation (FMSDF) state. A schematic diagram of the energy of fluctuations in this case is shown in Fig.~\ref{fig:03}. According to Eq.~\eqref{eq:18}, the critical temperature to support FMSDF, $T_{\text{FMSDF}}=1/(k_B\beta_{\text{FMSDF}})$, is determined by
\begin{equation}
    \frac{2\delta E}{g_{c}} = \frac{2}{3}\frac{\sinh{\beta_{\text{FMSDF}}\delta E}}{1+\cosh{\beta_{\text{FMSDF}}\delta E}}.
\end{equation}
The results are summarized in Fig.~\ref{fig:02}. 

\par Note that a nonzero quantum geometry of the flat band is necessary for the formation of FMSDF since the kinetic coefficient of the amplitude mode is proportional to the flat band's quantum weight, see Eq.~\eqref{eq:18}. In flat bands, the quenched dispersion suppresses kinetic energy, so the condensate naturally favors a spatially uniform OP. Yet, such a quench also removes the conventional stiffness arising from band dispersion that stabilizes uniformity, making the system prone to inhomogeneity. The quantum geometry, as an intrinsic inhomogeneity, not only introduces an unconventional stiffness governed by the quantum metric but also enhances the proneness toward inhomogeneity, thereby driving FMSDF.

\paragraph{Exciton-induced magnetism---}Applying an infinitesimal external in-plane Zeeman field $\mathbf{H}$ leads to the following source term in the effective action:
\begin{equation}\label{eq:22}
    S_{\text{eff}}^{(1)} \propto \sum_{\mathit{q}} \bigg( \mathbf{H}^{+}_{\mathit{q}}\Delta_{\mathit{q}} + \mathbf{H}^{-}_{\mathit{q}}\Delta^*_{\mathit{q}} \bigg),
\end{equation}
where $\mathbf{H}^{\pm}=(\mathbf{H}\cdot\mathbf{\hat{x}})\pm i(\mathbf{H}\cdot\mathbf{\hat{y}})$;  see SM for details \cite{SM}. Accordingly, the OP of the EI phases is proportional to the in-plane magnetization:
\begin{equation}\label{eq:23}
    \mathbf{M}(\mathbf{r}) \propto \text{Re}[\Delta(\mathbf{r})]\mathbf{\hat{x}} - \text{Im}[\Delta(\mathbf{r})]\mathbf{\hat{y}}.
\end{equation}
This indicates an exciton-induced in-plane paramagnetism due to the $U(1)$-symmetry of the free energy. Since the amplitude and phase modes correspond to fluctuations in the magnitude and direction of in-plane magnetization, the FMSDF state is characterized by a periodic modulation of the amplitude fluctuations. This leads to a susceptibility of the form
\begin{equation}
\langle \|\delta\mathbf{M}_{\mathbf{q}}\| \|\delta\mathbf{M}_{\mathbf{q}}\| \rangle \sim \frac{1}{\lambda_+(\mathbf{q})} \propto \frac{1}{A(T)-B_{ij}(T)\mathbf{q}^i\mathbf{q}^j},
\end{equation}
where $A(T), B_{ij}(T)>0$ are some temperature-dependent functions according to Eq.~\eqref{eq:18}. Identifying this form of susceptibility near $\mathbf{q}\to0$ when $T<T_{\text{FMSDF}}$ offers a potential route for detecting the FMSDF state.

\par Although our framework assumes negligible SOC, it is instructive to consider the case where SOC is much weaker than the out-of-plane Zeeman splitting and can be treated perturbatively, leading to the energy hierarchy
\begin{equation}\label{eq:21}
    g_{c} > 2\delta E \gg \lambda_{\text{SOC}},
\end{equation}
where $\lambda_{\text{SOC}}$ denotes the SOC strength. This introduces a linear term in the effective action, thereby favoring a net in-plane magnetization; see SM for details~\cite{SM}. This reveals that the $U(1)$ symmetry of the free energy stems from the spin-$U(1)$ symmetry of the underlying system.

\paragraph{Discussion and Materials Realization---}
Our theoretical framework is rigorously grounded under the energetic hierarchy $g_{c} > 2\delta E \gg \lambda_{\text{SOC}}$, where $\lambda_{\text{SOC}}$ is the characteristic spin-orbit coupling. A nearly vanishing SOC is required to ensure that the up and down spin bands share identical Bloch structures, thus maximizing the wavefunction overlap and isolating the pure geometrical effect. While perturbative inclusion of SOC would introduce weak in-plane anisotropy \cite{SM}, it does not erase the robust geometric scaling we report.

Covalent-organic frameworks (COFs) provide an ideal material platform to realize the criteria discussed above. Owing to their light constituent organic elements, COFs naturally host highly suppressed SOC \cite{doi:10.1021/acs.chemmater.1c04317, C8TA02555J}. Furthermore, their low dielectric screening drastically enhances local Coulomb interactions \cite{Shi2020, torabi2015}, readily satisfying the strong coupling condition ($g_{c} > 2\delta E$) necessary to overcome the Zeeman gap and induce condensation. Most importantly, the macroscopic lattice geometries of COFs, such as kagome and honeycomb-kagome architectures, can be precisely tailored to enforce intrinsic flat bands \cite{doi:10.1021/acs.accounts.0c00652, Jiang2019, D3SC06367D}, making them highly tunable templates for quantum geometry-driven correlated states. 

In conclusion, we have uncovered a fundamental mechanism wherein the quantum metric of a lattice-induced flat band not only stabilizes excitonic condensation but drives an anomalous, finite-momentum fluctuation state. By bridging flat-band quantum geometry with macroscopically observable magnetic fluctuations, this framework extends the frontier of strongly correlated phases in artificial lattices and highlights COFs as a promising playground for exploring the consequences of quantum geometry in materials.
\paragraph*{Acknowledgement---} This work was supported by the National Science Foundation through the Expand-QISE award NSF-OMA-2329067 and benefited from the resources of Northeastern University’s Advanced Scientific Computation Center, the Explorer Cluster, and the Massachusetts Technology Collaborative award MTC-22032.
\paragraph*{Data availability---}The data that support the findings of this article are not publicly available upon publication because it is not technically feasible and/or the cost of preparing, depositing, and hosting the data would be prohibitive within the terms of this research project. The data are available from the authors upon reasonable request.


\bibliography{apssamp}
\setcounter{equation}{0}
\setcounter{figure}{0}
\setcounter{table}{0}

\renewcommand{\theequation}{S\arabic{equation}}
\renewcommand{\thefigure}{S\arabic{figure}}
\renewcommand{\thetable}{S\arabic{table}}
\renewcommand{\bibnumfmt}[1]{[S#1]}
\renewcommand{\citenumfont}[1]{S#1}
\newcommand{\bk}{\boldsymbol\kappa}

\newcommand{\beginsupplement}{%
  \setcounter{equation}{0}
  \renewcommand{\theequation}{S\arabic{equation}}%
  \setcounter{table}{0}
  \renewcommand{\thetable}{S\arabic{table}}%
  \setcounter{figure}{0}
  \renewcommand{\thefigure}{S\arabic{figure}}%
  \setcounter{section}{0}
  \renewcommand{\thesection}{S\Roman{section}}%
  \setcounter{subsection}{0}
  \renewcommand{\thesubsection}{S\Roman{section}.\Alph{subsection}}%
}

\clearpage
\pagebreak
\begin{widetext}
\begin{center}
\textbf{\large Supplemental Material: Quantum Geometry-Driven Finite-Momentum Exciton Fluctuations in Flat-Band Systems}
\end{center}
\tableofcontents
\end{widetext}

\section{S1. Derivation of the effective action}\label{appx:A}
\par We derive the effective action for the order parameter (OP) shown in Eq.~\eqref{eq:action_eff}. We begin with the action for the interaction term of the flat-band electrons in Eq.~\eqref{eq:03}:
\begin{equation}
    S_{\text{int}} [\psi,\bar{\psi}] = -U_{c}\int^\beta_0 d\tau \int d[\mathbf{r}] \, \hat{b}^\dagger(\mathbf{r})\hat{b}(\mathbf{r}).
\end{equation}
Note that we have ignored terms renormalizing the chemical potential. We insert the following identity into the partition function \cite{PhysRevB.107.155142, PhysRevLett.132.026002, altland2010condensed, fluctuations_in_SC, Coleman_2015}:
\begin{equation}
    1 = \int \mathcal{D}[\xi,\xi^*]e^{-\frac{1}{U_{c}}\int^\beta_0 d\tau\int d[\mathbf{r}] \xi^*(\mathbf{r})\xi(\mathbf{r})},
\end{equation}
where the auxiliary field $\xi$ is further shifted into
\begin{equation}
    \xi(\mathbf{r}) = \Delta(\mathbf{r}) + U_{c}\hat{b}(\mathbf{r}).
\end{equation}
Since $\xi$ is a Gaussian field, it implies that $\langle\Delta(\mathbf{r})\rangle=U_{c}\langle \hat{b}(\mathbf{r})\rangle$, where the bracket refers to evaluating the expectation value. The partition function after the insertion becomes:
\begin{align}
    \mathcal{Z}= & \int \mathcal{D}[\Delta,\Delta^*] e^{-\frac{1}{U_{c}}\int^\beta_0 d\tau\int d[\mathbf{r}] \Delta^*(\mathbf{r})\Delta(\mathbf{r})} \notag
    \\ & \int \mathcal{D}[\psi,\bar{\psi}] e^{-S_0[\psi,\bar{\psi}]}e^{-S_{\text{mix}}[\psi,\bar{\psi},\Delta,\Delta^*]}.
\end{align}
Here, 
\begin{align}
    S_0[\psi,\bar{\psi}] & = \int^\beta_0 d\tau\int d[\mathbf{r}] \bar{\psi}_{\mathbf{r}} (\partial_\tau+\hat{H}_0(\mathbf{r})) \psi_{\mathbf{r}},
    \\ S_{\text{mix}}[\psi,\bar{\psi},\Delta,\Delta^*] & = \int^\beta_0 d\tau\int d[\mathbf{r}] \big( \Delta(\mathbf{r})\hat{b}^\dagger(\mathbf{r}) + \Delta^*(\mathbf{r})\hat{b}(\mathbf{r}) \big). 
\end{align}
We further write $S_{\text{mix}}[\psi,\bar{\psi},\Delta,\Delta^*]$ in the momentum space:
\begin{align}
        & S_{\text{mix}}[\psi,\bar{\psi},\Delta,\Delta^*] = \int^\beta_0 d\tau\int d[\mathbf{r}] \big( \Delta(\mathbf{r})\hat{b}^\dagger(\mathbf{r}) + \Delta^*(\mathbf{r})\hat{b}(\mathbf{r}) \big) \notag
        \\ &  = \int^\beta_0 d\tau\int d[\mathbf{r}] \int d[\mathbf{k}]d[\mathbf{p}]d[\mathbf{q}] \varphi^*_{\mathbf{p}}(\mathbf{r})\varphi_{\mathbf{q}}(\mathbf{r})\Delta_{\mathbf{k}}\hat{\gamma}^\dagger_{\uparrow,\mathbf{p}}\hat{\gamma}_{\downarrow,\mathbf{q}}e^{i\mathbf{k}\cdot\mathbf{r}}+ \text{H.c.} \notag
        \\ & = \int^\beta_0 d\tau \int d[\mathbf{k}]d[\mathbf{q}]\Lambda(\mathbf{q},-\mathbf{k})\Delta_\mathbf{k}\hat{\gamma}^\dagger_{\uparrow, \mathbf{k}+\mathbf{q}}\hat{\gamma}_{\downarrow,\mathbf{q}}+ \text{H.c.}, \label{eq:S7}
\end{align}
where we adapt the convention
\begin{equation}
    \Delta(\mathbf{r}) = \int d[\mathbf{k}]\Delta_\mathbf{k}e^{i\mathbf{k}\cdot\mathbf{r}}.
\end{equation}
The form factor $\Lambda(\mathbf{k},\mathbf{q})$ of the flat band is defined as \cite{PhysRevLett.132.026002}:
\begin{equation}
    \Lambda(\mathbf{k},\mathbf{q}) = \int d[\mathbf{r}]e^{-i\mathbf{q}\cdot\mathbf{r}}\varphi_{\mathbf{k-q}}^*(\mathbf{r})\varphi_{\mathbf{k}}(\mathbf{r}),
\end{equation}
which can be rewritten as:
\begin{equation}
    \Lambda(\mathbf{k},\mathbf{q}) = \braket{u_\mathbf{k-q}|u_\mathbf{k}},
\end{equation}
where we have expressed the cell-periodic part of the flat band's Bloch function $u_{\mathbf{k}}(\mathbf{r})=e^{-i\mathbf{k}\cdot\mathbf{r}}\varphi_{\mathbf{k}}(\mathbf{r})$ by the bra-ket notation $\ket{u_\mathbf{k}}$. We note that since $\langle\Delta_{\mathbf{k}}\rangle=U_{c}\langle \hat{b}_{\mathbf{k}}\rangle=U_{c}\int d[\mathbf{p}]\braket{u_\mathbf{p}|u_{\mathbf{p}+\mathbf{k}}}\langle\hat{\psi}^\dagger_{\downarrow,\mathbf{p}}\hat{\psi}_{\uparrow,\mathbf{p}+\mathbf{k}}\rangle$, the index in $\Delta_\mathbf{k}$ reflects the center-of-mass momentum of the exciton.

\par We can further rewrite Eq.~\eqref{eq:S7} as
\begin{equation}
    S_{\text{mix}}[\psi,\bar{\psi},\Delta,\Delta^*] = \int^\beta_0 d\tau\int d[\mathbf{k}]d[\mathbf{k}'] \bar{\psi}_{\mathbf{k}}\hat{H}_{\text{mix}}(\mathbf{k},\mathbf{k}')\psi_{\mathbf{k}'},
\end{equation}
where
\begin{equation}
\hat{H}_{\text{mix}}(\mathbf{k},\mathbf{k}') = \braket{u_\mathbf{k}|u_{\mathbf{k}'}}\begin{pmatrix} 0 & \Delta_{\mathbf{k}-\mathbf{k}'} \\ \Delta^*_{\mathbf{k}'-\mathbf{k}} & 0 \end{pmatrix}.
\end{equation}
Then, after integrating out the Grassmannian field $\psi$ \cite{fluctuations_in_SC, altland2010condensed, Coleman_2015}, the partition function becomes
\begin{equation}
    \mathcal{Z}=\int \mathcal{D}[\Delta,\Delta^*] e^{-\frac{1}{U_{c}}\int d\tau\int d[\mathbf{k}] \Delta^*_{\mathbf{k}}\Delta_{\mathbf{k}}} \text{det}(\hat{G}^{-1}),
\end{equation}
where
\begin{equation}
\begin{split}
    & \hat{G}^{-1}(\mathbf{k}, \mathbf{k}') =  \hat{G}_0^{-1}(\mathbf{k}, \mathbf{k}') + \hat{H}_{\text{mix}}(\mathbf{k},\mathbf{k}').
\end{split}
\end{equation}
and
\begin{align}
    \hat{G}_0^{-1}(\mathbf{k}, \mathbf{k}') & = \begin{pmatrix}  \partial_\tau+\delta E & 0 \\ 0 & \partial_\tau-\delta E\end{pmatrix}(2\pi)^2\delta(\mathbf{k}-\mathbf{k}'),
    \\ \hat{H}_{\text{mix}}(\mathbf{k},\mathbf{k}') & = \braket{u_\mathbf{k}|u_{\mathbf{k}'}}\begin{pmatrix} 0 & \Delta_{\mathbf{k}-\mathbf{k}'} \\ \Delta^*_{\mathbf{k}'-\mathbf{k}} & 0 \end{pmatrix}.
\end{align}
The effective action for the OP $S_{\text{eff}}[\Delta,\Delta^*]$ is
\begin{equation}\label{eq:S14}
\begin{split}
    & S_{\text{eff}}[\Delta,\Delta^*] = -\ln(\mathcal{Z})
    \\ & = \frac{1}{U_{c}}\int d\tau\int d[\mathbf{k}] \Delta^*_{\mathbf{k}}\Delta_{\mathbf{k}}  - \text{Tr}[\ln(1+\hat{G}_0\hat{H}_{\text{mix}})] 
    \\ & = \frac{1}{U_{c}}\int d\tau\int d[\mathbf{k}] \Delta^*_{\mathbf{k}}\Delta_{\mathbf{k}} + \sum_{m=1}^{\infty}\frac{1}{m}\text{Tr}\big[ (-\hat{G}_0\hat{H}_{\text{mix}})^m \big],
\end{split}
\end{equation}
Expanding the second term in the last line to order $\mathcal{O}(|\Delta|^4)$ restores Eq.~\eqref{eq:action_eff}.

\par Note that since $\hat{H}_{\text{mix}}$ contains only off-diagonal matrix elements and $\hat{G}_0$ contains only diagonal matrix elements, there are only $\mathcal{O}(|\Delta|^2)$ and $\mathcal{O}(|\Delta|^4)$ terms in the effective action upon expanding the second term in Eq.~\eqref{eq:S14} to order  $\mathcal{O}(|\Delta|^4)$, denoting as $S_{\text{eff}}^{(2)}$ and $S_{\text{eff}}^{(4)}$, respectively. To effectively compute $S_{\text{eff}}^{(2)}$ and $S_{\text{eff}}^{(4)}$, we work with Matsubara frequency representations, using the decomposition \cite{PhysRevB.107.155142}:
\begin{equation*}
    \Delta_{\mathbf{q}}(\tau)=\sqrt{\frac{1}{\beta}}\sum_{\omega_n}\Delta_{(\mathbf{q},i\omega_n)}e^{-i\omega_n\tau}, 
\end{equation*}
where $\omega_n$ denotes the bosonic Matsubara frequency. Then, $S_{\text{eff}}^{(2)}$ is \cite{altland2010condensed}:
\begin{align}
    S_{\text{eff}}^{(2)} & = \frac{1}{U_{c}}\sum_{\mathit{q}}\Delta_{\mathit{q}}^*\Delta_{\mathit{q}} + \frac{1}{2}\text{Tr} \big[ (-\hat{G}_0\hat{H}_{\text{mix}})^2] \notag
    \\ & = \sum_{\mathit{q}} \big[ \frac{1}{U_{c}} + \frac{1}{\beta}\sum_{\mathit{k}}|\braket{u_{\mathbf{k+q}}|u_{\mathbf{k}}}|^2G_{11}(\mathit{k})G_{22}(\mathit{k+q}) \big] \Delta_{\mathit{q}}^*\Delta_{\mathit{q}}. \label{eq:S15}
\end{align}
Here, $\mathit{k}=(\mathbf{k},i\epsilon_n)$ and $\mathit{q}=(\mathbf{q},i\omega_n)$, $\sum_{\mathit{q}}\equiv \sum_{i\omega_n}\int d[\mathbf{q}]$, and
\begin{align}
    G_{11}(\mathit{k}) & = \frac{1}{i\epsilon_n-\delta E}
    \\ G_{22}(\mathit{k}) & = \frac{1}{i\epsilon_n+\delta E},
\end{align}
where $\epsilon_n$ denotes the fermionic Matsubara frequency. $S_{\text{eff}}^{(4)}$ can be obtained with the same techniques:
\begin{widetext}
\begin{align}
    S_{\text{eff}}^{(4)} & = \frac{1}{4}\text{Tr} \big[ (-\hat{G}_0\hat{H}_{\text{mix}})^4] \notag
    \\ & = \frac{1}{2\beta^2}\sum_{\mathit{k}_1,\mathit{k}_3,\mathit{k}_5,\mathit{k}_7} \big[ \braket{u_{\mathbf{k}_3}|u_{\mathbf{k}_1}}\braket{u_{\mathbf{k}_5}|u_{\mathbf{k}_3}}\braket{u_{\mathbf{k}_7}|u_{\mathbf{k}_5}}\braket{u_{\mathbf{k}_1}|u_{\mathbf{k}_7}}
     G_{11}(\mathit{k}_1)G_{11}(\mathit{k}_5)G_{22}(\mathit{k}_3)G_{22}(\mathit{k}_7)\Delta_{\mathit{k}_3-\mathit{k}_1}\Delta_{\mathit{k}_3-\mathit{k}_5}^*\Delta_{\mathit{k}_7-\mathit{k}_5}\Delta_{\mathit{k}_7-\mathit{k}_1}^* \big]. \label{eq:S18}
\end{align}
\end{widetext}
Then, we define $\mathit{q}_1=\mathit{k}_7-\mathit{k}_1$, $\mathit{q}_2=\mathit{k}_3-\mathit{k}_5$, $\mathit{q}_3=\mathit{k}_3-\mathit{k}_1$, $\mathit{q}_4=\mathit{k}_7-\mathit{k}_5$, and $\mathit{q}=(\mathit{k}_1+\mathit{k}_3)/2$. Equation~\eqref{eq:S18} becomes:
\begin{widetext}
\begin{equation}\label{eq:S19}
\begin{split}
     S_{\text{eff}}^{(4)} & = \frac{1}{2\beta^2}\sum_{\mathit{q}_1,\mathit{q}_2,\mathit{q}_3,\mathit{q}_4,\mathit{q}}\braket{u_{\mathbf{q}+\frac{\mathbf{q}_3}{2}}|u_{\mathbf{q}-\frac{\mathbf{q}_3}{2}}}\braket{u_{\mathbf{q}+\frac{\mathbf{q}_3}{2}-\mathbf{q}_2}|u_{\mathbf{q}+\frac{\mathbf{q}_3}{2}}}\braket{u_{\mathbf{q}+\frac{\mathbf{q}_1-\mathbf{q}_2+\mathbf{q}_4}{2}}|u_{\mathbf{q}+\frac{\mathbf{q}_1-\mathbf{q}_2-\mathbf{q}_4}{2}}}\braket{u_{\mathbf{q}+\frac{-\mathbf{q}_1-\mathbf{q}_2+\mathbf{q}_4}{2}}|u_{\mathbf{q}+\frac{\mathbf{q}_1-\mathbf{q}_2+\mathbf{q}_4}{2}}}
    \\ & \quad \quad \quad \quad \quad \quad \quad \times G_{11}(\mathit{q}-\frac{\mathit{q}_3}{2})G_{11}(\frac{\mathit{q}_1-\mathit{q}_2-\mathit{q}_4}{2}+\mathit{q})G_{22}(\mathit{q}+\frac{\mathit{q}_3}{2})G_{22}(\frac{\mathit{q}_1-\mathit{q}_2+\mathit{q}_4}{2}+\mathit{q})\Delta_{\mathit{q}_1}^*\Delta_{\mathit{q}_2}^*\Delta_{\mathit{q}_3}\Delta_{\mathit{q}_4}\delta_{\mathit{q}_1+\mathit{q}_2,\mathit{q}_3+\mathit{q}_4}.
\end{split}
\end{equation}
\end{widetext}
Here, $\delta_{q_1,q_2}=(2\pi)^2\delta_{\omega_{n,1},\omega_{n,2}}\delta(\mathbf{q}_1-\mathbf{q}_2)$. According to Eqs.~\eqref{eq:S15} and \eqref{eq:S19}, we can get the coefficients in Eq.~\eqref{eq:action_eff} as:
\newpage
\begin{widetext}
\begin{align}
    a(\mathit{q}) & = \frac{1}{U_{c}} + \frac{1}{\beta}\sum_{\mathit{k}}|\braket{u_{\mathbf{k+q}}|u_{\mathbf{k}}}|^2G_{11}(\mathit{k})G_{22}(\mathit{k+q}),  \label{eq:S20}
    \\ b(\{\mathit{q}_n\}) & = \frac{1}{2\beta^2}\sum_{\mathit{q}}\braket{u_{\mathbf{q}+\frac{\mathbf{q}_3}{2}}|u_{\mathbf{q}-\frac{\mathbf{q}_3}{2}}}\braket{u_{\mathbf{q}+\frac{\mathbf{q}_3}{2}-\mathbf{q}_2}|u_{\mathbf{q}+\frac{\mathbf{q}_3}{2}}}\braket{u_{\mathbf{q}+\frac{\mathbf{q}_1-\mathbf{q}_2+\mathbf{q}_4}{2}}|u_{\mathbf{q}+\frac{\mathbf{q}_1-\mathbf{q}_2-\mathbf{q}_4}{2}}}\braket{u_{\mathbf{q}+\frac{-\mathbf{q}_1-\mathbf{q}_2+\mathbf{q}_4}{2}}|u_{\mathbf{q}+\frac{\mathbf{q}_1-\mathbf{q}_2+\mathbf{q}_4}{2}}} \notag
    \\ & \quad \quad \quad \quad \quad \quad \times G_{11}(\mathit{q}-\frac{\mathit{q}_3}{2})G_{11}(\frac{\mathit{q}_1-\mathit{q}_2-\mathit{q}_4}{2}+\mathit{q})G_{22}(\mathit{q}+\frac{\mathit{q}_3}{2})G_{22}(\frac{\mathit{q}_1-\mathit{q}_2+\mathit{q}_4}{2}+\mathit{q})\delta_{\mathit{q}_1+\mathit{q}_2,\mathit{q}_3+\mathit{q}_4}. \label{eq:S21}
\end{align}
\end{widetext}

\section{S2. Derivation of the mean-field theory and Gaussian fluctuations}\label{appx:B}
\par We now show that the coefficients in the effective action connect to the quantum metric of the flat band in a certain limit and derive effective actions with mean-field approximation to study phase transitions and fluctuations around the mean-field solution. First, we note that for the long-wavelength modes such that $\mathbf{q}\to0$, $a(\mathit{q})$ in Eq.~\eqref{eq:S20} becomes:
\begin{widetext}
\begin{align}
    \lim_{\mathbf{q}\to0} a(\mathit{q}) & = \frac{1}{U_{c}} + \frac{1}{\beta}\int d[\mathbf{k}]\lim_{\mathbf{q}\to0}|\braket{u_{\mathbf{k+q}}|u_{\mathbf{k}}}|^2\sum_{\epsilon_n}G_{11}(\epsilon_n)G_{22}(\epsilon_n+\omega_n) \notag
    \\ & = \frac{1}{U_{c}} + \frac{1}{\beta}\int d[\mathbf{k}]\lim_{\mathbf{q}\to0}(1-g_{ij}(\mathbf{k})\mathbf{q}^i\mathbf{q}^j)\sum_{\epsilon_n}G_{11}(\epsilon_n)G_{22}(\epsilon_n+\omega_n) \notag
    \\ & = \frac{1}{U_{c}} + \frac{\Omega_{\text{BZ}}}{\beta }\sum_{\epsilon_n}G_{11}(\epsilon_n)G_{22}(\epsilon_n+\omega_n) - \lim_{\mathbf{q}\to0}\frac{1}{\beta}\sum_{\epsilon_n}G_{11}(\epsilon_n)G_{22}(\epsilon_n+\omega_n)\frac{K_{ij}}{2\pi}\mathbf{q}^i\mathbf{q}^j \notag
    \\ & = \frac{1}{U_{c}} + \Omega_{\text{BZ}}\frac{f_D(\delta E)-f_D(-\delta E)}{i\omega_n+2\delta E} - \lim_{\mathbf{q}\to0}\frac{f_D(\delta E)-f_D(-\delta E)}{i\omega_n+2\delta E}\frac{K_{ij}}{2\pi}\mathbf{q}^i\mathbf{q}^j.
\end{align}
\end{widetext}
Here, $\Omega_{\text{BZ}}\equiv\int d[\mathbf{k}]1$ is the area of BZ, $f_D(E)$ is the Fermi-Dirac distribution, $g_{ij}(\mathbf{k})=\braket{\partial_{k^i}u_{\mathbf{k}}|(1-\ket{u_{\mathbf{k}}}\bra{u_{\mathbf{k}}})|\partial_{k^j}u_{\mathbf{k}}}$ is the quantum metric, and $K_{ij}=2\pi\int d[\mathbf{k}]g_{ij}(\mathbf{k})$ is the quantum weight \cite{PhysRevX.14.011052, PhysRevLett.133.206602, PhysRevResearch.7.023158}. We have used the identity $\lim_{\mathbf{q}\to0}|\braket{u_{\mathbf{k+q}}|u_{\mathbf{k}}}|^2=\lim_{\mathbf{q}\to0}(1-g_{ij}(\mathbf{k})\mathbf{q}^i\mathbf{q}^j)$, which can be directly shown as follows
\begin{widetext}
\begin{align}
    \lim_{\mathbf{q}\to0}|\braket{u_{\mathbf{k+q}}|u_{\mathbf{k}}}|^2 \approx & \lim_{\mathbf{q}\to0}|(\bra{u_{\mathbf{k}}}+\mathbf{q}^i\bra{\partial_{k^i}u_{\mathbf{k}}} + \frac{\mathbf{q}^i\mathbf{q}^j}{2}\bra{\partial_{k^ik^j}u_{\mathbf{k}}}+\mathcal{O}(|\mathbf{q}|^3))\ket{u_{\mathbf{k}}}|^2, \notag
    \\ \approx &  \lim_{\mathbf{q}\to0} 1 + \mathbf{q}^i\mathbf{q}^j\left[ \braket{\partial_i u_{\mathbf{k}}|u_{\mathbf{k}}}\braket{u_{\mathbf{k}}|\partial_j u_{\mathbf{k}}} + \frac{1}{2}( \braket{\partial_{k^ik^j}u_{\mathbf{k}}|u_{\mathbf{k}}} + \braket{u_{\mathbf{k}}|\partial_{k^ik^j}u_{\mathbf{k}}} )\right] +\mathcal{O}(|\mathbf{q}|^3) \notag
    \\ = & \lim_{\mathbf{q}\to0} 1 + \mathbf{q}^i\mathbf{q}^j\left[ \braket{\partial_i u_{\mathbf{k}}|u_{\mathbf{k}}}\braket{u_{\mathbf{k}}|\partial_j u_{\mathbf{k}}} - \braket{\partial_{k^i}u_{\mathbf{k}}|\partial_{k^j}u_{\mathbf{k}}} \right] \notag
    \\ = & \lim_{\mathbf{q}\to0} 1 - g_{ij}(\mathbf{k})\mathbf{q}^i\mathbf{q}^j.
\end{align}
\end{widetext}

\par Considering the static limit such  that $\omega\to0$ following the analytical continuation $i\omega_n\to\omega+i0^+$ \cite{PhysRevB.107.155142}, we derive:
\begin{equation}
    a(\mathit{q}) = \frac{\sinh{\beta\delta E}}{1+\cosh{\beta\delta E}}\frac{K_{ij}}{4\pi\delta E}\mathbf{q}^i\mathbf{q}^j + \frac{1}{U_{c}} -  \frac{\Omega_{\text{BZ}}\sinh{\beta\delta E}}{2\delta E(1+\cosh{\beta\delta E})},
\end{equation}
which restores Eq.~\eqref{eq:09}. Accordingly, the coherence length is \cite{grosso2013solid}:
\begin{equation}
    \xi_{\text{c}}^2 = \frac{K_{0}}{4\pi \Omega_{\text{BZ}}}|\frac{2\delta E(1+\cosh{\beta\delta E})}{g_{c}\sinh{\beta\delta E}} -  1|^{-1},
\end{equation}
which restores Eq.~\eqref{eq:xi_c}. Here, we define $g_{c}\equiv \Omega_{\text{BZ}}U_{c}$.

\par We proceed to derive the effective actions within the mean-field approximation, and separate the order parameter into \cite{PhysRevB.107.155142}:
\begin{equation}\label{eq:S24}
    \Delta_{\mathit{q}} = \sqrt{\beta}\Delta_0\delta_{\mathit{q},0} + \delta\Delta_{\mathit{q}}.
\end{equation}
Here, $\Delta_0$ denotes the mean-field value of the order parameter, and $\delta \Delta_{\mathit{q}}$ describes the fluctuation near $\mathit{q}=0$. Similarly, we decompose the effective action into:
\begin{equation}
    S_{\text{eff}} = S_0 + S_{\text{G}},
\end{equation}
in which we consider terms only to order $\mathcal{O}(|\delta \Delta|^2)$ in $S_{\text{G}}$. Specifically,
\begin{equation}
    S_0 = a_0|\Delta_0|^2 + b_0|\Delta_0|^4,
\end{equation}
where 
\begin{align}
    a_0 & = \Omega_{\text{BZ}}\beta\bigg[ \frac{1}{g_{c}} - \frac{\sinh{\beta\delta E}}{2\delta E(1+\cosh{\beta\delta E})} \bigg] \notag
    \\ & = \frac{\beta \Omega_{\text{BZ}}}{2\delta E}(\frac{2\delta E}{g_{c}} - \frac{\sinh{\beta\delta E}}{1+\cosh{\beta\delta E}}), \label{eq:S27}
    \\ b_0 & = \frac{\Omega_{\text{BZ}}}{2}\sum_{\epsilon_n}[G_{11}(\epsilon_n)]^2[G_{22}(\epsilon_n)]^2 \notag
    \\ & = \frac{\beta \Omega_{\text{BZ}}}{8(\delta E)^3}\frac{\sinh{\beta\delta E}}{(1+\cosh{\beta\delta E})},
\end{align}
which restore Eqs.~\eqref{eq:13} and \eqref{eq:14}. By defining the dimensionless temperature $\tilde{\beta}\equiv\beta\delta E$ and the dimensionless interaction strength $\tilde{g}\equiv g_{c}/2\delta E$, the mean-field critical temperature $T_{\text{MF}}$ in units of $\delta E/k_B$ as a function of $2\delta E/g_{c}$ is the positive solution of the equation
\begin{equation}\label{eq:SCF_MF}
    \frac{\sinh{\tilde{\beta}}}{1+\cosh{\tilde{\beta}}}-\frac{1}{\tilde{g}}=0,
\end{equation}
which indicates that $g_{c}>2\delta E$ is necessary to have a phase transition.

\par The effective action for the fluctuation $S_{\text{G}}$ can be obtained by substituting Eq.~\eqref{eq:S24} into Eqs.~\eqref{eq:S15} and \eqref{eq:S19} and keeping terms only to order $\mathcal{O}(|\delta\Delta|^2)$, leading to:
\begin{widetext}
\begin{equation*}
\begin{split}
    S_{\text{G}} = & \sum_{\mathit{p}} \big[ \frac{1}{U_{c}} + \frac{1}{\beta}\sum_{\mathit{q}}|\braket{u_{\mathbf{q+p}}|u_{\mathbf{q}}}|^2G_{11}(\mathit{q})G_{22}(\mathit{q+p}) \big] \delta\Delta_{\mathit{p}}^*\delta\Delta_{\mathit{p}}
    \\ & +\frac{1}{\beta}\sum_{\mathit{p},\mathit{q}}|\braket{u_{\mathbf{q+p}}|u_{\mathbf{q}}}|^2G_{11}(\mathit{q})G_{11}(\mathit{q}+\mathit{p})(G_{22}(\mathit{q}+\mathit{p}))^2|\Delta_0|^2\delta\Delta_{\mathit{p}}^*\delta\Delta_{\mathit{p}}
    \\ & +\frac{1}{2\beta}\sum_{\mathit{p},\mathit{q}}|\braket{u_{\mathbf{q+p}}|u_{\mathbf{q}}}|^2G_{11}(\mathit{q})G_{11}(\mathit{p}+\mathit{q})G_{22}(\mathit{q})G_{22}(\mathit{q}+\mathit{p})(\Delta_0^*)^2\delta\Delta_{\mathit{p}}\delta\Delta_{-\mathit{p}}
    \\ & +\frac{1}{2\beta}\sum_{\mathit{p},\mathit{q}}|\braket{u_{\mathbf{q+p}}|u_{\mathbf{q}}}|^2G_{11}(\mathit{q})G_{11}(\mathit{p}+\mathit{q})G_{22}(\mathit{q})G_{22}(\mathit{p}+\mathit{q})(\Delta_0)^2\delta\Delta_{\mathit{p}}^*\delta\Delta_{-\mathit{p}}^*
    \\ & +\frac{1}{\beta}\sum_{\mathit{p},\mathit{q}}|\braket{u_{\mathbf{q+p}}|u_{\mathbf{q}}}|^2(G_{11}(\mathit{q}))^2G_{22}(\mathit{q})G_{22}(\mathit{p}+\mathit{q})|\Delta_0|^2\delta\Delta_{\mathit{p}}^*\delta\Delta_{\mathit{p}},
\end{split}
\end{equation*}
\end{widetext}
which can be written in the compact form:
\begin{equation}
    S_{\text{G}} = \frac{1}{2}\sum_{\mathit{p}} \begin{pmatrix} \delta\Delta^*_{\mathit{p}} & \delta\Delta_{-\mathit{p}} \end{pmatrix} \begin{pmatrix} \alpha_1(\mathit{p},|\Delta_0|) & \alpha_2(\mathit{p},\Delta_0) \\ \alpha_2^*(\mathit{p},\Delta_0) & \alpha_1(\mathit{p},|\Delta_0|) \end{pmatrix} \begin{pmatrix} \delta\Delta_{\mathit{p}} \\ \delta\Delta^*_{-\mathit{p}}\end{pmatrix},
\end{equation}
where $\mathit{p}=(\mathbf{p},i\omega_n)$ and
\begin{widetext}
\begin{align}
    \alpha_1(\mathit{p},|\Delta_0|)  = & \frac{1}{U_{c}} + \big[ -\frac{1}{i\omega_n+2\delta E} + |\Delta_0|^2\frac{i\omega_n+4\delta E}{2(\delta E)^2(i\omega_n+2\delta E)^2} \big]\frac{\sinh{\beta\delta E}}{1+\cosh{\beta\delta E}}\int d[\mathbf{q}]|\braket{u_{\mathbf{q+p}}|u_{\mathbf{q}}}|^2 \label{eq:S31}
    \\ \alpha_2(\mathit{p},\Delta_0) = & \frac{(\Delta_0)^2}{\delta E(\omega_n^2+4(\delta E)^2)}\frac{\sinh{\beta\delta E}}{1+\cosh{\beta\delta E}}\int d[\mathbf{q}]|\braket{u_{\mathbf{q+p}}|u_{\mathbf{q}}}|^2. \label{eq:S32}
\end{align}
\end{widetext}
We can further diagonalize $S_{\text{G}}$ in the basis of $(\eta_{+,\mathit{p}},\,\eta_{-,\mathit{p}})$, which relates to $\delta \Delta_{\mathit{p}}$ by:
\begin{equation}
    \eta_{\pm,\mathit{p}} = \frac{\sqrt{\beta}}{2}\bigg( e^{i\theta_0}\delta\Delta_{\mathit{p}} \pm e^{-i\theta_0}\delta\Delta^*_{-\mathit{p}} \bigg),
\end{equation}
where $\theta_0=\text{Arg}(\Delta_0)$. Then, $S_{\text{G}}$ becomes:
\begin{equation}
    S_{\text{G}} = \sum_{\xi=\pm}\sum_{\mathit{p}}\lambda_{\xi}(\mathit{p},|\Delta_0|)\eta_{\xi,\mathit{p}}^*\eta_{\xi,\mathit{p}},
\end{equation}
where
\begin{equation}\label{eq:S35}
\begin{split}
    \lambda_{\pm}(\mathit{p},|\Delta_0|) & = 2\beta[ \alpha_1(\mathit{p},|\Delta_0|) \pm |\alpha_2(\mathit{p},\Delta_0)|].
\end{split}
\end{equation}
Since $\eta_{\pm,-\mathit{p}}^* = \pm\eta_{\pm,\mathit{p}}$, the introduced fluctuations $\eta_+$ and $\eta_-$ are real and imaginary in real space, corresponding to the amplitude and phase mode, respectively \cite{nagaosa1999quantum}. 

\par By substituting Eqs.~\eqref{eq:S31} and \eqref{eq:S32} into \eqref{eq:S35} and taking the limit $\mathit{p}\to0$, Eqs. \eqref{eq:17} and \eqref{eq:18} can be restored. Specifically, when $\mathbf{p},\omega\to0$, Eqs.~\eqref{eq:S31} and \eqref{eq:S32} become:
\begin{widetext}
\begin{align}
    \alpha_1(\mathit{p},|\Delta_0|) & \cong \frac{1}{U_{c}} + \big[ -\frac{1}{2\delta E} + |\Delta_0|^2\frac{1}{2(\delta E)^3} \big]\frac{\sinh{\beta\delta E}}{1+\cosh{\beta\delta E}} (\Omega_{\text{BZ}}-\frac{K_{ij}\mathbf{p}^i\mathbf{p}^j}{2\pi}) \notag
    \\ & = -\frac{1}{\beta}a_0 + \big[ \frac{1}{2\delta E}\frac{\sinh{\beta\delta E}}{1+\cosh{\beta\delta E}} + \frac{2}{\beta \Omega_{\text{BZ}}}a_0 \big]\frac{K_{ij}\mathbf{p}^i\mathbf{p}^j}{2\pi} \notag
    \\ & = -\frac{1}{\beta}a_0(1-\frac{K_{ij}\mathbf{p}^i\mathbf{p}^j}{2\pi \Omega_{\text{BZ}}}) + \big[ \frac{\Omega_{\text{BZ}}}{2\delta E}\frac{\sinh{\beta\delta E}}{1+\cosh{\beta\delta E}} + \frac{1}{\beta}a_0 \big]\frac{K_{ij}\mathbf{p}^i\mathbf{p}^j}{2\pi \Omega_{\text{BZ}}} \label{eq:S36}
    \\ |\alpha_2(\mathit{p},\Delta_0)| & \cong \frac{|\Delta_0|^2}{4(\delta E)^3}\frac{\sinh{\beta\delta E}}{1+\cosh{\beta\delta E}}(\Omega_{\text{BZ}}-\frac{K_{ij}\mathbf{p}^i\mathbf{p}^j}{2\pi}) \notag
    \\ & = \frac{2}{\beta}|\Delta_0|^2b_0(1-\frac{K_{ij}\mathbf{p}^i\mathbf{p}^j}{2\pi \Omega_{\text{BZ}}}) \notag
    \\ & = -\frac{1}{\beta}a_0(1-\frac{K_{ij}\mathbf{p}^i\mathbf{p}^j}{2\pi \Omega_{\text{BZ}}}) \label{eq:S37}
    ,
\end{align}
\end{widetext}
where we have substituted $|\Delta_0|^2$ by the value $-a_0/2b_0$. Then,
\begin{widetext}
\begin{align}
    \lambda_{+}(\mathbf{p}) & = -4a_0(1-\frac{K_{ij}\mathbf{p}^i\mathbf{p}^j}{2\pi \Omega_{\text{BZ}}}) + 2\beta\big[ \frac{\Omega_{\text{BZ}}}{2\delta E}\frac{\sinh{\beta\delta E}}{1+\cosh{\beta\delta E}} + \frac{1}{\beta}a_0 \big]\frac{K_{ij}\mathbf{p}^i\mathbf{p}^j}{2\pi \Omega_{\text{BZ}}} \notag
    \\ & =  -4a_0 + 2\beta\big[ \frac{\Omega_{\text{BZ}}}{2\delta E}\frac{\sinh{\beta\delta E}}{1+\cosh{\beta\delta E}} + \frac{3}{\beta}a_0 \big]\frac{K_{ij}\mathbf{p}^i\mathbf{p}^j}{2\pi \Omega_{\text{BZ}}} \notag
    \\ & = \frac{2\beta \Omega_{\text{BZ}}}{\delta E}\bigg( \frac{\sinh{\beta\delta E}}{1+\cosh{\beta\delta E}} -\frac{2\delta E}{g_{c}} \bigg) +\frac{3\beta}{\delta E} \bigg(\frac{2\delta E}{g_{c}} - \frac{2}{3}\frac{\sinh{\beta\delta E}}{1+\cosh{\beta\delta E}} \bigg) \frac{K_{ij}}{2\pi}\mathbf{p}^i\mathbf{p}^j \label{eq:S38}
    \\ \lambda_{-}(\mathbf{p}) & = 2\beta\big[ \frac{1}{2\delta E}\frac{\sinh{\beta\delta E}}{1+\cosh{\beta\delta E}} + \frac{1}{\beta}a_0 \big]\frac{K_{ij}\mathbf{p}^i\mathbf{p}^j}{2\pi \Omega_{\text{BZ}}}  \notag
    \\ & = \frac{\beta}{g_{c}}\frac{K_{ij}}{\pi}\mathbf{p}^i\mathbf{p}^j. \label{eq:S39}
\end{align}
\end{widetext}
Note that we have used Eq.~\eqref{eq:S27} in Eq.~\eqref{eq:S39}. According to Eq.~\eqref{eq:S38}, the kinetic coefficient of the amplitude mode is positive only when:
\begin{equation}
    \frac{3}{2}\frac{2\delta E}{g_{c}} > \frac{\sinh{\beta\delta E}}{1+\cosh{\beta\delta E}} > \frac{2\delta E}{g_{c}},
\end{equation}
where the first and second inequalities refer to a positive kinetic coefficient and the establishment of mean-field order, respectively. Since $\sinh{x}/(1+\cosh{x})<1$ for finite $x$, a negative kinetic coefficient may occur at finite temperature if the interaction is too strong such that $2\delta E/g_{c}<2/3$. This suggests that the amplitude mode with finite momentum is favored for strong enough interaction. However, it does not develop a density wave order. Note that since the quantum metric term $(\Omega_{\text{BZ}}-\frac{K_{ij}}{2\pi}\mathbf{p}^i\mathbf{p}^j)$ originates from the expansion of $\int d[\mathbf{k}]|\braket{u_{\mathbf{k+p}}|u_{\mathbf{k}}}|^2$, the momentum cutoff must be chosen such that $(\Omega_{\text{BZ}}-\frac{K_{ij}}{2\pi}\mathbf{p}^i\mathbf{p}^j)\geq0$. Accordingly, when the kinetic coefficient is negative, the amplitude mode is bounded from below within the momentum cutoff:
\begin{equation}
    \lambda_{+}(\mathbf{p}) \geq \frac{2\beta}{g_{c}}.
\end{equation}
Therefore, we term such a state a finite-momentum superfluid density fluctuation as it does not alter the ground state nor develop a quasi-long-range order for a finite interaction strength. 

\section{S3. Calculation of the Berezinskii–Kosterlitz–Thouless (BKT) transition temperature}\label{appx:C}
To solve Eq.~\eqref{eq:19} self-consistently, we first write down $|\Delta_{0,\text{BKT}}|^2$ explicitly:
\begin{align}
    |\Delta_{0,\text{BKT}}|^2 & = 2(\delta E)^2\bigg(1 -\frac{(1+\cosh{\tilde{\beta}})}{\tilde{g}\sinh{\tilde{\beta}}} \bigg),
\end{align}
where $\tilde{g}\equiv g_{c}/2\delta E$ and $\tilde{\beta}\equiv \beta_{\text{BKT}}\delta E$ are the dimensionless interaction strength and dimensionless temperature, respectively. Then, we can rewrite Eq.~\eqref{eq:17} as
\begin{align}
    & \frac{1}{\beta_{\text{BKT}}} = |\Delta_{0,\text{BKT}}|^2\frac{1}{g_{c}}\frac{K_0}{4} \notag
    \\ \rightarrow & \frac{1}{\beta_{\text{BKT}}} = 2(\delta E)^2\bigg(1 -\frac{(1+\cosh{\tilde{\beta}})}{\tilde{g}\sinh{\tilde{\beta}}} \bigg)\frac{1}{g_{c}}\frac{K_0}{4} \notag
    \\ \rightarrow & 1=\frac{\tilde{\beta}}{\tilde{g}}\bigg(1 -\frac{(1+\cosh{\tilde{\beta}})}{\tilde{g}\sinh{\tilde{\beta}}} \bigg)\frac{K_0}{4}.
\end{align}
We further define the function $f_{\tilde{g},K_0}(\tilde{\beta})$ for a given $\tilde{g}$ and $K_0$ as:
\begin{equation}
    f_{\tilde{g},K_0}(\tilde{\beta})\equiv \frac{\tilde{\beta}}{\tilde{g}}\bigg(1 -\frac{(1+\cosh{\tilde{\beta}})}{\tilde{g}\sinh{\tilde{\beta}}} \bigg)\frac{K_0}{4} - 1.
\end{equation}
The BKT temperature is the solution of $f_{\tilde{g},K_0}(\tilde{\beta})=0$ that are greater than $\delta E/(k_BT_{\text{MF}})$.

\section{S4. Connection to in-plane magnetization and the effect of spin-orbit coupling (SOC)}\label{appx:D}
\par The OP fluctuations connect to the fluctuations of in-plane magnetization once a spatially varying in-plane magnetic field is introduced as a perturbation such that its strength is much smaller than the Zeeman field. The action for applying the spatially varying in-plane magnetic field $\mathbf{H}(\mathbf{r})=\int d[\mathbf{k}]\mathbf{H}_\mathbf{k}e^{i\mathbf{k}\cdot\mathbf{r}}$ to the system is:
\begin{equation}
\begin{split}
    & S_{\mathbf{H}}[\psi,\bar{\psi}] = \int^\beta_0 d\tau\int d[\mathbf{r}] \big( \mathbf{H}^{+}(\mathbf{r})\hat{b}^\dagger(\mathbf{r}) + \mathbf{H}^{-}(\mathbf{r})\hat{b}(\mathbf{r}) \big)
    \\ & = \int^\beta_0 d\tau\int d[\mathbf{k}]d[\mathbf{k}'] \bar{\psi}_{\mathbf{k}} \braket{u_\mathbf{k}|u_{\mathbf{k}'}}\begin{pmatrix} 0 & \mathbf{H}^{-}_{\mathbf{k}-\mathbf{k}'} \\ \mathbf{H}^{+}_{\mathbf{k}'-\mathbf{k}} & 0 \end{pmatrix}\psi_{\mathbf{k}'},
\end{split}
\end{equation}
where $\mathbf{H}^{\pm}=(\mathbf{H}\cdot\mathbf{\hat{x}})\pm i(\mathbf{H}\cdot\mathbf{\hat{y}})$. Then, the effective action (and hence, the free energy) will have terms linear in the OP:
\begin{align}\label{eq:S58}
    S_{\text{eff}}^{(1)} = & \sum_{\mathit{q}} \bigg[ \frac{1}{U_{c}} + \frac{1}{\beta}\sum_{\mathit{k}}|\braket{u_{\mathbf{k+q}}|u_{\mathbf{k}}}|^2G_{11}(\mathit{k})G_{22}(\mathit{k+q}) \bigg] \notag
    \\ & \times \bigg( \mathbf{H}^{+}_{\mathit{q}}\Delta_{\mathit{q}} + \mathbf{H}^{-}_{\mathit{q}}\Delta^*_{\mathit{q}} \bigg). 
\end{align}
But, the free energy of magnetization $\mathbf{M}$ coupled to an external field $\mathbf{H}$ is:
\begin{equation}
    F = -\int d[\mathbf{r}] \mathbf{M}(\mathbf{r})\cdot\mathbf{H}(\mathbf{r}).
\end{equation}
Therefore, the in-plane magnetization is proportional to the OP:
\begin{align}
    \mathbf{M}(\mathbf{r}) \propto & -\frac{1}{\beta}\bigg( \frac{1}{g_{c}} -  \frac{\sinh{\beta\delta E}}{2\delta E(1+\cosh{\beta\delta E})} \bigg) \notag
    \\ & \quad \times \bigg( \text{Re}[\Delta(\mathbf{r})]\mathbf{\hat{x}} - \text{Im}[\Delta(\mathbf{r})]\mathbf{\hat{y}} \bigg), \label{eq:S60}
\end{align}
where we only consider the constant terms in the coefficient in Eq.~\eqref{eq:S58}. According to Eq.~\eqref{eq:S60}, the amplitude mode $\eta_+$ corresponds to fluctuations in the magnitude of the in-plane magnetization, and the phase mode $\eta_-$ corresponds to fluctuations in the direction of the in-plane magnetization.

\par Note that the SOC integrates into the effective interaction akin to the in-plane magnetic field, provided its magnitude remains significantly smaller than that of the Zeeman field, allowing it to be regarded as a perturbative term. Such a perturbative SOC adds a linear component to the effective action similar to that found in Eq.~\eqref{eq:S58}. This explicitly disrupts the $U(1)$-symmetry of the free energy, thereby promoting the exciton condensate with a specific phase and, consequently, favoring a certain direction for the corresponding in-plane magnetization.


\end{document}